\documentclass[paper]{JHEP3} % 10pt is ignored!

\usepackage{epsfig,multicol,amsmath,amsfonts,amssymb,cite}

\newcommand\fverb{\setbox\pippobox=\hbox\bgroup\verb}
\newcommand\fverbdo{\egroup\medskip\noindent%
                        \fbox{\unhbox\pippobox}\ }
\newcommand\fverbit{\egroup\item[\fbox{\unhbox\pippobox}]}
\newbox\pippobox

\title{Matching Regge Theory to the OPE}

\author{ S.S.~Afonin$^{\ast}$,  A.A.~Andrianov$^{\ast}$, 
V.A.~Andrianov$^{\ast}$~and  D.~Espriu$^{\diamondsuit}$\\
 $^{\ast}$V.A.~Fock Department of Theoretical Physics,
St.~Petersburg State University,\\ 1 ul.Ulianovskaya, 198504, Russia\\
$^{\diamondsuit}$Departament d'Estructura i Constituents de la
Mat\`eria and
 CER for Astrophysics, Particle Physics and Cosmology,
 Universitat de Barcelona, 647 Diagonal, 08028, Spain\\
E-mail: \email{afonin24@mail.ru}, 
\email{andrianov@bo.infn.it}, \email{Vladimir.Andrianov@pobox.spbu.ru}, 
\email{espriu@ecm.ub.es}}

\preprint{March 2004\\UB-ECM-PF-04/06}

\abstract{
The spectra of masses and decay constants for non-strange meson resonances 
in the energy range 0--2.5 GeV is analyzed. It is known from meson 
phenomenology that for given quantum numbers these spectra approximately 
follow linear trajectories with a universal slope. These facts can be 
understood in terms of an effective string description for QCD. For light 
meson states the trajectories deviate noticeably from the linear behavior. 
We investigate the possible corrections to the linear trajectories by
matching two-point correlators of quark currents to the Operator Product 
Expansion (OPE). We find that the allowed modifications to the linear 
Regge behavior must decrease rapidly with the principal quantum number. After
fitting the lightest states in each channel and certain low-energy constants
the whole spectrum for meson masses and residues is obtained in a satisfactory 
agreement with phenomenology. We briefly speculate on possible implications 
for the QCD effective string.}


\begin{document}

\section{Introduction}

Hadron phenomenology tells us that the masses
squared of mesons with given quantum numbers lie approximately
on equally-spaced
linear trajectories\footnote{From now on, we shall term this behavior
"linear trajectory ansatz" (LTA).}
(see, for example, the modern
reviews~\cite{ani, ani2}). This is a strong indication that QCD admits
an effective string description, as this type of spectrum is
characteristic e.g. of the bosonic\footnote{This string model is of course
not consistent, a more realistic amplitude is the Lovelace-Shapiro
amplitude but unfortunately it cannot be derived from any known string
and
as we shall see is also incompatible with QCD anyway.
See e.g. \cite{AABE} for a discussion on this point.}
string\cite{ven}.
Long ago it was noticed\cite{bgs} that
if one accepts this behavior, by including an infinite number
of resonances it is possible to reproduce
the parton-model logarithm present in two-point correlators.

This observation means that the exchange of an infinite number
of vector mesons may be dual to the perturbative
QCD continuum made of quarks and gluons. This quark-hadron duality was
explicitly verified in QCD in 1+1
dimensions in the large-$N_c$ limit~\cite{hoof,two}. At present, the
QCD community understand by quark-hadron duality several, somewhat different,
 things.
The most extended assertion is that
perturbation theory can be used to calculate a certain smeared
hadronic cross section (so-called global duality)~\cite{PQW}.
We use the term in a rather lax sense and understand by quark-hadron
duality any matching of hadronic and
partonic degrees of freedom.

In the context of large-$N_c$ QCD, quark-hadron duality for mesons
has received a lot of attention
\cite{kat1}--\cite{PhLB}.
In\cite{kat1} the parameters of linear vector mass spectrum were
derived within the framework of finite-energy sum rules.
The axial-vector channel was also considered there. Later, this
approach was extended to the pseudoscalar\cite{kat2} and
scalar\cite{kat3} cases.
In~\cite{gesh} the problem of matching this infinite set of meson states
to the Operator Product Expansion (OPE)\cite{svz}
was considered, with $m^2_V(n)$ being a general
function. In the review~\cite{sh}
possible deviations from quark-hadron duality were investigated (including
meson resonance widths, i.e. away from the large-$N_c$ limit.)
In~\cite{p1} the linear trajectory
ansatz in the vector and axial channels were used  to
calculate certain physical observables. Subsequently, in
\cite{beane}--\cite{PhLB} various aspects of
quark-hadron duality, chiral symmetry restoration (CSR), and
matching to the OPE have been discussed.

Since, after all, it is well known that the linear trajectory ansatz fails
to describe the lightest states in each channel (particularly so in the
pseudoscalar channel),
some attempts to go
beyond the linear trajectory ansatz were studied in~\cite{beane2,sim},
in the latter paper using a
semi-classical string analysis. The ans$\ddot{\text{a}}$tze proposed
there, however, seem to lead to a discrepancy 
with the OPE that will be explained later.

In the present work we want to re-take these issues. Our interest has,
to some extent, been triggered by the recent controversy among the authors
of references \cite{beane} and \cite{p3}, but we shall also report here on
some older results of us\cite{garda}.
We shall consider radial Regge trajectories for meson resonances with the same 
quantum numbers (so-called ``radial excitations'') in
the vector~($V$), axial-vector~($A$), scalar~($S$), and pseudoscalar~($P$)
correlation functions.
If one believes
in the existence of a QCD effective string --- and we do ---
the slope of trajectories must be the same in all channels
because this slope is proportional
to the string tension that depends on gluodynamics only
(string universality). In this respect we tend to agree with the approach of
\cite{beane}, but we find other aspects of this work to be more questionable.

We propose a systematic method to take into account
possible corrections to the linear trajectories in the $V,A,S,P$
channels. Our ansatz for the meson mass spectra, once substituted
into two-point correlators, is matched to
the OPE, which gives a set of constraints on meson mass parameters.
As we shall see, the OPE dictates a very particular class of corrections
to the linear trajectory ansatz. These corrections are not anecdotic: they
turn out to be essential to get good phenomenology and by taking them into
account one does not need
to introduce an artificial separation between low-lying and excited radial
excitations in the different channels like in
\cite{beane,p1,p2,p3}.
Finally, the results are fitted to the experimental
data~\cite{ani,ani2,pdg}.

The paper is organized as follows. In section~2 we remind the
general idea of how to construct the asymptotic sum rules for the
case of infinite number of resonances. In particular, the status
and the necessity of non-linear corrections to the linear trajectory
ansatz is thoroughly examined in sections~3 and 4 for the case of $V,A$
mesons. In section~5 this analysis is extended to the $S,P$ channels.
Section~6
contains the details of our fits and discussion of the obtained
results. In section~7 we outline our conclusions.

\section{Current-current correlators and scheme of sum rules}

Let us consider the two-point correlators of $V,A,S,P$ quark currents
in the large-$N_c$ limit of QCD. On the one hand, by virtue of
confinement they are saturated by an infinite set of narrow meson
resonances, that is, they can be  represented by the sum
of narrow meson resonances with given quantum numbers $J$ and masses
$m_J (n),\, n =0,1,\ldots$,

\begin{equation}
\label{cor1}
\Pi^J(Q^2)=\int d^4x\exp(iQx)\langle\bar{q}\Gamma q(x)\bar{q}\Gamma q(0)
\rangle_{\text{planar}}=
\sum_n\frac{Z_J(n)}{Q^2+m_J^2(n)}+D_0^J+D_1^JQ^2,
\end{equation}
\begin{equation}
J\equiv S,P,V,A; \qquad
\Gamma=i,\gamma_{5},\gamma_{\mu},\gamma_{\mu}\gamma_{5}; \qquad
D_{0},D_{1}=const.
\end{equation}
The last two terms both in the $S,P$ and in the
$V,A$ channels represent a perturbative contribution,
with $D_0$ and $D_1$ being contact terms required to eliminate
divergences in the ultraviolet. On the other hand,
their high-energy asymptotics is
provided~\cite{svz,rry} by perturbation theory and the OPE. In the chiral
limit\footnote{In~\cite{z1} asymptotic terms governed by 
the dimension-two gluon condensate $\lambda^2$ 
("gluon mass") were
introduced. This dimension-two condensate cannot be produced by 
a local
gauge-invariant operator~\cite{z2,kondo}. On the other hand, 
this condensate brings  little 
influence on the fits of meson parameters (see the discussion in Appendix A).
We put here this condensate to zero.}
\begin{equation}
\label{V}
\Pi^V(Q^2)=\frac{1}{4\pi^2}\left(1+\frac{\alpha_s}{\pi}\right)
\ln\!\frac{\mu^2}{Q^2}
+\frac{\alpha_s}{12\pi}\cdot
\frac{\langle(G_{\mu\nu}^a)^2\rangle}{Q^4}-\frac{28}{9}\pi\alpha_s
\frac{\langle\bar{q}q\rangle^2}{Q^6}\,,%+\mathcal{O}\left(\frac{1}{Q^8}\right),
\end{equation}
\begin{equation}
\label{A}
\Pi^A(Q^2)=\frac{1}{4\pi^2}\left(1+\frac{\alpha_s}{\pi}\right)
\ln\!\frac{\mu^2}{Q^2}
+\frac{\alpha_s}{12\pi}\cdot
\frac{\langle(G_{\mu\nu}^a)^2\rangle}{Q^4}+\frac{44}{9}\pi\alpha_s
\frac{\langle\bar{q}q\rangle^2}{Q^6}\,,%+\mathcal{O}\left(\frac{1}{Q^8}\right),
\end{equation}
\begin{equation}
\label{S}
\Pi^S(Q^2)=-\frac{3}{8\pi^2}\left(1+\frac{11\alpha_s}{3\pi}\right)Q^2
\ln\!\frac{\mu^2}{Q^2}
+\frac{\alpha_s}{8\pi}\cdot
\frac{\langle(G_{\mu\nu}^a)^2\rangle}{Q^2}-\frac{22}{3}\pi\alpha_s
\frac{\langle\bar{q}q\rangle^2}{Q^4}\,,%+\mathcal{O}\left(\frac{1}{Q^6}\right),
\end{equation}
\begin{equation}
\label{P}
\Pi^P(Q^2)=-\frac{3}{8\pi^2}\left(1+\frac{11\alpha_s}{3\pi}\right)Q^2
\ln\!\frac{\mu^2}{Q^2}
+\frac{\alpha_s}{8\pi}\cdot
\frac{\langle(G_{\mu\nu}^a)^2\rangle}{Q^2}+\frac{14}{3}\pi\alpha_s
\frac{\langle\bar{q}q\rangle^2}{Q^4}\,,%+\mathcal{O}\left(\frac{1}{Q^6}\right),
\end{equation}
where we have defined
\begin{equation}
\label{trans}
\Pi_{\mu\nu}^{V,A}(Q^2)\equiv\left(-\delta_{\mu\nu}Q^2+Q_{\mu}Q_{\nu}\right)
\Pi^{V,A}(Q^2)\,.
\end{equation}
The vacuum saturation for the 4-fermion condensate\cite{svz} in the
large-$N_c$ limit has been used. In Eqs.~(\ref{V})-(\ref{P})
 $\langle(G_{\mu\nu}^a)^2\rangle $ and the $\langle\bar{q}q\rangle $
represent the gluon and the quark condensate, respectively.
The normalization scale $\mu$ arises after additive renormalization
of the infinite sums in  Eq.~(\ref{cor1}).  $\alpha_s$ is the QCD
coupling constant taken at the Chiral Symmetry Breaking (CSB)
scale $\sim 1$ GeV.

As it can be seen from Eqs.~(\ref{V})-(\ref{P}) differences of
correlators of opposite parity  decrease very rapidly at large momenta
\begin{align}
\label{VA}
\Pi^V(Q^2)-\Pi^A(Q^2)&=-8\pi\alpha_s\frac{\langle\bar
qq\rangle^2}{Q^6}+\mathcal{O}\left(\frac{1}{Q^8}\right),\\
\Pi^S(Q^2)-\Pi^P(Q^2)&=-12\pi\alpha_s\frac{\langle\bar
qq\rangle^2}{Q^4}+\mathcal{O}\left(\frac{1}{Q^6}\right).
\end{align}
This signifies the chiral symmetry restoration (CSR) at
high energies, with the quark condensate~$\langle\bar{q}q\rangle$
being an order parameter of the chiral symmetry breaking in QCD.

Expanding the right-hand side of Eq.~(\ref{cor1}) in inverse powers
of $Q^2$ and comparing
with Eqs.~(\ref{V})-(\ref{P}) one obtains several asymptotic 
sum rules for each channel
(we do not account for condensates of dimension eight and higher). If one
deals with a finite number of resonances the procedure is
straightforward (for the CSR sum rules\cite{aet, aaa}) . In the case of infinite number of resonances and
given an ansatz $m^2(n)$ one uses
the Euler-Maclaurin summation formula or, alternatively, the $\psi$-function
(both methods are equivalent). It should be noticed here that,
generally speaking, applying a summation procedure for a divergent sum and
expanding in $1/Q^2$ are not commutative operations. A regulator is
in general
called for.

The sum rules, obtained in this way, deliver certain constraints on
the parameters of meson mass spectra. This information can be
extracted and compared with phenomenology, which is the subject
of the next sections.

In this work we shall ignore the anomalous dimensions of the operators
and the running of the coupling constant, and therefore we do not take
into account the logarithmic corrections that they may introduce in the
sum rules. We shall come back to this issue in a separate paper.

\section{Linear trajectories for vector and axial-vector resonances}

In the $V,A$ case the residues $Z_{V,A}(n)$ in Eq.~(\ref{cor1}) have the
structure
\begin{equation}
\label{VAres}
Z_{V,A}(n)=2F_{V,A}^2(n),
\end{equation}
where the quantities $F_{V,A}(n)$ represent the decay constants
parameterizing the matrix elements of  the corresponding currents
\begin{align}
\label{vme}
\langle0|j_{\text{em}}^{\mu}(0)|V(\epsilon,k,n)
\rangle=\frac{1}{(2\pi)^{\frac32}}
eF_V(n)m_V(n)\epsilon^{\mu},\\
\langle0|A^{\mu}(0)|A(\epsilon,k,n)\rangle=\frac{1}{(2\pi)^{\frac32}}
eF_A(n)m_A(n)\epsilon^{\mu}.
\label{ame}
\end{align}
where $k$ is the meson momentum, $\epsilon^\mu$ is a polarization
vector, and $e$ is electron charge.
The quantity $F_{V}(n)$ can be obtained from the decay of vector mesons
\begin{equation}
\label{vres}
\Gamma_{V\rightarrow e^+e^-}(n)=\frac{4\pi\alpha^2F_{V}^2(n)}{3m_{V}(n)},
\end{equation}
where $\alpha$ is the fine structure constant.
The constants $F_{A}(n)$ are related to the widths of some radiative decays.
Unfortunately, this sector is poorly known, except for the reactions
including the ground axial-vector state ($a_1$-meson). We have found two
examples\cite{isg,P}
\begin{equation}
\Gamma_{\tau\rightarrow a_1\nu_{\tau}}=\frac{G_F^2m_{\tau}^3F_{a_1}^2}
{16\pi}\left(1-\frac{m_{a_1}^2}{m_{\tau}^2}\right)
\left(1+\frac{2m_{a_1}^2}{m_{\tau}^2}\right),
\end{equation}
and
\begin{equation}
\Gamma_{a_1\rightarrow\pi\gamma}=\frac{\alpha
F_{a_1}^2m_{a_1}}
{24f_{\pi}^2}\left(1-\frac{m_{\pi}^2}{m_{a_1}^2}\right)^3.
\end{equation}
where $G_F$ is the Fermi constant, $m_{\tau}$ is the mass of the
$\tau$-lepton, and $f_{\pi}$ is the pion decay constant.

Let us now introduce
the linear trajectory ansatz for the meson mass spectrum
\begin{equation}
\label{LTA}
m_{V,A}^2(n)=M^2_{V,A}+a_{V,A}n\,;\qquad F_{V,A}^2(n)=const\equiv
F^2_{V,A}; \qquad n=0,1,2,\dots\,,
\end{equation}
with constant intercepts $M^2_{V,A}$ and slopes $a_{V,A}$.
In order to do the  sums over all resonances we shall use
the Euler-Maclaurin summation formula in Eq.~(\ref{cor1})
$$
\sum_{n=0}^Nf(n)=\int_0^{N}\!\!f(x)\,dx+\frac12[f(0)+f(N)]+\frac{B_1}{2!}
[f'(N)-f'(0)]-
$$
\begin{equation}
\label{EM}
\frac{B_2}{4!}[f'''(N)-f'''(0)]+ .\,.\,. +
(-1)^k\frac{B_{k+1}}{(2k+2)!}[f^{(2k+1)}(N)-f^{(2k+1)}(0)]+\dots\,,
\end{equation}
where $B_1=1/6\,,B_2=1/30\,,\dots$ are Bernoulli numbers.
We shall expand the final expression in powers of $Q^{-2}$ and after comparing
with the OPE's \eqref{V} and \eqref{A} one gets the set of asymptotic  sum
rules.

At this point we have to introduce a regulator\footnote{Another way to
improve the convergence is to differentiate with respect to $Q^2$. This
recipe will be used below for the matching to QCD.}
since the series is manifestly
not absolutely convergent (the general term behaves like $\sim 1/n $). We
shall cut-off the infinite sum by including only the $N$ first terms in the
sum. This naturally cuts off the restricted value of momenta to
$ \Lambda_{\mbox{\scriptsize{cut};V,A}}^2\equiv 
M^2_{V,A}+a_{V,A} N_{V,A} \simeq a_{V,A} N_{V,A}$. 
As we identify parity-doublers
between the $V$- and $A$-channels and do not intent the cutoff 
to break chiral symmetry, the cutoff numbers $N_V$ and $N_A$ 
must be taken equal, 
$N_V = N_A$. On the other hand, the appropriate cutoffs in the OPE also 
coincide in the $V,A$-channels in a chirally 
symmetric regularization\footnote{Note that a momentum
cut-off is questionable in the
OPE expansions on account of gauge invariance , 
but perfectly gauge invariant in the sum over resonances
\eqref{cor1}. So one is actually forced to use different regulators on
both sides, but after the renormalization with the help of constants
$D_0^{V,A}$
the leading logarithmic term in $\mu^2$ is
in any case unambiguous.}, so
$ \Lambda_{\mbox{\scriptsize{cut};V}}= \Lambda_{\mbox{\scriptsize{cut};A}}$ 
(see also similar arguments in \cite{beane2}). 
Therefore for a 
natural implementation of chiral
symmetry the slopes of radial Regge trajectories must be the same, $a_V = a_A$.
To compare with OPE~(\ref{V}) and~(\ref{A}) one has
to make an additive renormalization by means of subtraction of
infinite constants $D_0^{V,A}$. In order not to break the chiral
symmetry the subtraction must be equal for parity doublers,
$D_0^V=D_0^A$. The conclusion is then
\begin{equation}
\label{sr1}
\frac43\cdot\frac{N_c}{16\pi^2}=\frac{2F_V^2}{a_V}=\frac{2F_A^2}{a_A}.
\end{equation}
We remark that the QCD
string model implies $a_V=a_A\equiv a$ as well. (This relation $a_V=a_A$ is not
fulfilled in~\cite{p1,p2} where such analysis for the LTA was performed.) Then
one concludes that $F_V=F_A$. Note that the limit $N\to \infty$, or
equivalently, $\Lambda_{\mbox{\scriptsize{cut}}} \to \infty$ is assumed. This
allows us to neglect terms of the form
$Q^2/\Lambda_{\mbox{\scriptsize{cut}}}^2$ or
$m^2/\Lambda_{\mbox{\scriptsize{cut}}}^2$.

However, as we have already mentioned, it is not possible to extract
further consequences from the individual $V$ and $A$ asymptotic sum rules.
Sub-leading terms are somewhat ambiguous. 
Fortunately, the annoying logarithm is
absent in differences of correlation functions. Namely
\begin{equation}
\label{V-A}
\Pi^V(Q^2)-\Pi^A(Q^2)=2\sum_{n=0}^{\infty}\left(\frac{F_V^2(n)}{Q^2+m_V^2(n)}-
\frac{F_A^2(n)}{Q^2+m_A^2(n)}\right)
\end{equation}
This difference (and the equivalent for $S,P$) show a very fast CSR. The
leading contribution is of ${\cal O}(1/Q^6)$: all perturbative
and purely gluonic contributions cancel in the difference. For
individual sum rules we shall perform the appropriate derivatives
in $Q^2$ before matching OPE and radial Regge trajectories.

The above fast convergence in correlator differences was used among others by \cite{beane},
where the following generalization of the Weinberg sum rules~\cite{weinberg}
was proposed
\begin{gather}
\label{bea2}
\sum_{n=0}^{\infty}\left(F_V^2(n)m_V^{2i}(n)-F_A^2(n)m_A^{2i}(n)\right)=
C^{(i)}, \qquad i=0,1, \\
C^{(0)}=f_{\pi}^2,\qquad C^{(1)}=0. \notag
\end{gather}
We emphasize that
one sums over chiral pairs in the difference (\ref{bea2}). This
becomes very
important at the moment that one needs to cut-off the sums with a finite
value of  $N$. If the cut-off is placed in such a way that the
``chiral partner'' of a given resonance, that is included, is left out, chiral symmetry will
be explicitly broken by the regulator.
It is thus vital to properly identify "chiral partners".
This is not an issue in the $V,A$ cases, but it will be so as we shall see
in the $S,P$ channels.

From the second sum rule ($i=1$) it immediately follows, given that
$F_A=F_V$ and $a_A=a_V$, that $M_{A}=M_{V}$. This automatically implies
if the believe in the LTA that the spectrum of QCD is degenerate in the
vector and axial-vector channels, something that it is obviously not true
(even after correcting for the quark mass differences, not included in the
present analysis). Furthermore, the first sum rule ($i=0$), implies
$f_\pi=0$, something again manifestly incompatible with experimental
evidence.
Since $f_\pi$ can be interpreted as the order parameter of chiral symmetry,
what we are seeing here is that the LTA corresponds to a string model that
has many common features with QCD, but an essential one is missing, namely
it would correspond to a string model without chiral symmetry
breaking\footnote{The Lovelace-Shapiro amplitude, after correcting the
intercept has the proper Adler zero, required by chiral
symmetry. However, it still has regularly spaced levels, something that
as we see is manifestly incompatible with QCD asymptotics. It cannot
be the correct QCD string. The interested reader may want to
see \cite{AABE} for an attempt to
formulate consistent string propagation in a background where chiral
symmetry has been broken.}.
Note that
convergence of the infinite sums in Eq.~\eqref{bea2} also requires to
impose the equality of slopes $a_V=a_A$, intercepts
$M_V=M_A$, and decay constants $F_V=F_A$.

Of course one may escape the contradiction between the
fact that the LTA leads to $f_\pi=0$ and the real world by declaring that
the ground states (perhaps even the first few resonances) in each channel,
e.g. the $\rho$ and $a_1$ mesons, are isolated resonances.
This pattern roughly meets the existing spectroscopy of $V,A$ mesons, but it
does not reproduce the physical values of condensates\cite{p1}.
However, such an ansatz satisfies the CSR~\eqref{bea2} at high energies.
Obviously this way out is quite ad-hoc and cannot be justified on any
QCD-inspired string model.

In the present paper we want to argue that the ans$\ddot{\text{a}}$tze
proposed in~\cite{p1,beane} can be improved by introducing certain systematic
corrections to the linear
trajectories~(\ref{LTA}). The problem here is that arbitrary
ans$\ddot{\text{a}}$tze for $m^2(n)$ and $F^2(n)$
result in appearance of terms which are absent in the standard OPE,
namely, terms with a fractional power of~$Q^2$ and terms of the kind
$Q^{-2k}\ln\!\dfrac{\Lambda_{\mbox{\scriptsize{cut}}}^2}{Q^2}$. (Recall
that we do not consider anomalous dimensions.)

\section{Sum rules and deviation from the LTA in the $V,A$ channels}

Our goal is to construct a class of radial Regge trajectories that does not
lead to the unwanted terms. Namely, our ansatz must reproduce
the parton-model logarithm and contain only  inverse powers
of large momentum squared $Q^2$ besides the logarithm. Recall that,
using (\ref{EM}),
\begin{equation}
\sum_{n=0}^N\frac{F^2(n)}{Q^2+m^2(n)}=\int_0^N\!\!
\frac{F^2(x)\,dx}{Q^2+m^2(x)}+\mathcal{O}\left(\frac{1}{Q^{2}}\right)\,.
\end{equation}
Clearly the logarithm can be only produced by the integral.
In order to generate the logarithm we require that
\begin{equation}
\label{plog}
\int\!\frac{F^2(x)\,dx}{Q^2+m^2(x)}+D_0=C\ln\!\left(\frac{Q^2+m^2(x)}{\mu^2}
\right)+ \mathcal{O}\left(\frac{1}{Q^{2}}\right),
\end{equation}
here $C$ is a constant from OPE (3) and (4), $D_0$ is a subtraction
constant, and $\mu$ is a normalization scale.

Differentiating Eq.~(\ref{plog}) with respect
to $x$, one can see that the two requirements are satisfied only if
\begin{equation}
\label{conVA}
F^2(x)=t\left(x\right)\frac{dm^2(x)}{dx}\,,\quad
t\left(x\right)=C+\Delta t\left(x\right)\,,
\end{equation}
where $\Delta t\left(x\right)$ is a decreasing function to be defined.
If we do not consider anomalous dimensions the function
$\Delta t\left(x\right)$ does not have to induce any
logarithms and other non-polynomial in $Q^{-2}$ terms in the 
integral~(\ref{plog}). Thus, the
direct expansion of the integral
$$
\int_0^{m^2(N)}\frac{\Delta t\left(x\right)d\left(m^2(x)\right)}
{Q^2+m^2(x)}=\sum_{k>0}t_k\left(\frac{1}{Q^2}\right)^k,
$$
where
\begin{equation}
\label{coef}
t_k=\int_0^{m^2(N)}\Delta t\left(x\right)\left(-m^2(x)\right)^{k-1}
d\left(m^2(x)\right),
\end{equation}
must exist at any order.

Convergence of the integrals in~\eqref{coef} for any $k$ is possible if the
function $\Delta t\left(x\right)$ falls off as an exponential, or faster.
Therefore, since one does not have yet any
dynamical arguments let us take the simplest possibility, namely
\begin{equation}
\label{t1}
\Delta t\left(x\right)=A_{F}e^{-B_{F}x}\,,\qquad
B_{F}>0\,,
\end{equation}
with constant $A_F$ and $B_F$.
Given an ansatz for $m^2(n)$, Eq.~(\ref{conVA}) together
with~(\ref{t1}) provides the condition of consistency with the OPE.
It is clear that the LTA ansatz~(\ref{LTA}) is a simple
particular case of Eq.~(\ref{conVA}). Any improvement of the
ansatz~(\ref{LTA}) has to meet Eq.~(\ref{conVA}) as well.
We remark that the corrections to linear trajectories, proposed
in~\cite{beane2,sim}, do not satisfy this requirement. Thus,
these ans$\ddot{\text{a}}$tze cannot be matched to the OPE.

The sum rules~\eqref{bea2} can actually be generalized to values of
$i$ greater than 1
\begin{equation}
\label{bea}
\sum_{n=0}^{\infty}\left(F_V^2(n)m_V^{2i}(n)-F_A^2(n)m_A^{2i}(n)\right)=
C^{(i)}, \qquad i=0,1,\dots\,.
\end{equation}
Now the $C^{(i)}$ contain a contribution from higher dimensional
condensates. For
the absolute convergence of Eq.~(\ref{bea}) at a given $i$ one has
to have
\begin{gather}
F_V^2(n)-F_A^2(n)\sim\frac{1}{n^{1+\alpha}}, \qquad \alpha>i,\\
m_V^2(n)-m_A^2(n)\sim\frac{1}{n^{\beta}}, \qquad \beta>i. \label{28}
\end{gather}

Let us now discuss corrections to Eq.~(\ref{LTA})
\begin{equation}
\label{acor1}
m^2_{V,A}(n)=M_{V,A}^2+a\,n+\delta_{V,A}(n)\,.
\end{equation}
Note that the convergence for any $i$ requires
$M_V=M_A\equiv M$ and that
the contribution from the function $\delta_{V,A}(n)$ to the 
difference~\eqref{28} must decrease at least exponentially.

Collecting our observations, we propose the simplest ansatz
satisfying all previous requirements, keeping of course in mind
that this is just one of the possibilities
\begin{align}
\label{VAmasses}
m^2_{V,A}(n)&=M^2+an+A_m^{V,A}e^{-B_m n},  \\
\label{acor2}
F^2_{V,A}(n)&=a\left(C+A_F^{V,A}e^{-B_F n}\right), \\
C&=\frac{1}{8\pi^2}\left(1+\frac{\alpha_s}{\pi}\right),\qquad
B_m>0, \qquad B_F>0,
\label{CVA}
\end{align}
with certain constants $A_{m,F}^{V,A}$, $B_{m,F}$ to be fitted.
We do not know the underlying dynamics responsible for the
appearance of those exponential corrections. But it seems for us
reasonable to suppose that, for masses, this dynamics is governed
mostly by gluons and thereby does not
depend on flavor. Thus, we keep the exponent $B_m$ the same for
all channels. For the same reason we regard $B_F$ as independent
of parity. We note also that in Eq.~\eqref{conVA} it is enough
to retain only the linear in~$n$ part of~$m^2(n)$.

In what follows we adopt a perturbative approach and
retain in the sum rules only terms linear in the
exponentially small corrections. Products of exponentials
exceed the precision of
our simple (one-exponential) parameterization: these
products are regarded as of the order of next-to-next-to-leading
corrections to Eqs.~\eqref{VAmasses} and~\eqref{acor2}.
One should keep 
in mind the precision of our approach and thus, when comparing the OPE
and the sums of resonances, 
 we only match the perturbative logarithms and the
first non-perturbative contribution $\mathcal{O}(1/Q^4)$  in the $V,A$ 
channels 
separately, and the leading
(non-perturbative) correction $\mathcal{O}(1/Q^6)$ in the difference
$\Pi^V(Q^2)-\Pi^A(Q^2)$ which represents a true order parameter of
the chiral symmetry breaking in the chiral limit.

In order to avoid of irrelevant infinite constant
let us consider the first derivative of $V,A$-correlators.
Introducing the notation for the linear part,
\begin{equation}
\label{lin}
\bar{m}^2(n)\equiv M^2+an,
\end{equation}
and making use of~\eqref{VAres} and~\eqref{VAmasses}-\eqref{CVA} we have
(the indices $V,A,$ are dropped for brevity)
\begin{equation}
\label{cor}
\frac{d\Pi(Q^2)}{dQ^2}=-2\sum_{n=0}^{\infty}
\frac{a\left(C+A_Fe^{-B_Fn}\right)}{\left(Q^2+\bar{m}^2(n)+
A_me^{-B_mn}\right)^2}.
\end{equation}
For the ground states the exponential corrections are not small
(in general, about 50\%). We will not apply our perturbative
approach for these mesons. Separating out the ground states and
retaining in the remainder only the parts linear 
in exponential corrections, one has
for~\eqref{cor}
\begin{multline}
\label{cor2}
-\frac12\frac{d\Pi(Q^2)}{dQ^2}\simeq
\frac{a(C+A_F)}{\left(Q^2+M^2+A_m\right)^2} \\
+\sum_{n=1}^{\infty}\left\{\frac{aC}{\left(Q^2+\bar{m}^2(n)\right)^2}+
\frac{aA_Fe^{-B_Fn}}{\left(Q^2+\bar{m}^2(n)\right)^2}-
\frac{2aCA_me^{-B_mn}}{\left(Q^2+\bar{m}^2(n)\right)^3}\right\}
\end{multline}
In the r.h.s. of Eq.~\eqref{cor2} one has three sums. The first
one represents the first derivative of $\psi$-function, which has the
standard asymptotic expansion at large $Q^2$. In the second and third sums
one may permute the summation over $n$ and expansion in $Q^{-2}$
due to absolute convergence in any order of this expansion. The
final expressions for the sum rules are presented in Appendices~B and C.
(see Eqs.~(\ref{qfourv})-(\ref{last})).

The solution of these equations will be considered after
discussion of the scalar sector. In addition, the so-called D-wave vector
mesons should be taken into account, see Appendix~D.

As a further evidence for the need of nonlinear corrections
let us consider the chiral symmetry restoration limit
$\langle\bar qq\rangle\rightarrow 0$ and $f_{\pi}\rightarrow 0$.
As follows from  Eqs.~\eqref{qfourv},\eqref{qfoura}, in the case of LTA
the corresponding solution $M^2=\frac12a$ gives a negative
value for the gluon condensate 
in Eqs.~\eqref{gv},~\eqref{ga}. This implies instability of
vacuum energy\cite{instab} and, hence, the appearance of tachyons. In that
sense the LTA is inconsistent with rather general principles, even in the
chiral symmetry restoration limit. A consistent ansatz can be only non-linear.

\section{Sum rules for scalar and pseudoscalar resonances}

The scalar case differs from the vector one and the derivation presented
in the previous section needs to be revisited.
Due to the transversality~\eqref{trans}, the
$V,A$ correlators required only one subtraction in Eq.~\eqref{cor1}.
The $S,P$ correlators require two subtractions.
Accepting the definition\footnote{This formula does not work for the
$\pi$-meson if we include the pion in the radial Regge trajectory. Thus for the
corresponding residue we just accept its value in the current algebra:
$Z_{\pi} = 2\frac{\langle\bar qq\rangle}{f_{\pi}^2}$. As we anyway separate
out the lowest state, this does not affect the subsequent analysis. 
Alternatively one could extend Eq.~\eqref{SPres}
with the help of a constant shift for all resonances in $Z_{S,P}(n)$
by a quantity $Z_{\pi}$. However this modification 
induces an unacceptably large value of dimension-two condensate 
in the OPE (see discussion in Appendix A). 
In accordance to  known 
theoretical and phenomenological
estimations \cite{z1,z2,domsch} we put this condensate to zero.
}
\begin{equation}
\label{SPres}
Z_{S,P}(n)\equiv2G_{S,P}^2(n)m_{S,P}^2(n),
\end{equation}
the analog
of~(\ref{plog}) in the $S,P$ case takes the form
\begin{equation}
\label{slog}
\int\frac{G^2(x)m^2(x)\,dx}{Q^2+m^2(x)}+D_0+D_1Q^2=
-\bar C Q^2\ln\!\left(\frac{Q^2+m^2(x)}{\mu^2}\right)+\dots\,,
\end{equation}
where $\bar{C}>0$, and $D_0,D_1$ are subtraction constants. Let us
make the following rearrangement. The sum over resonances
can be rewritten as
\begin{multline}
\label{ren}
\Pi(Q^2)= 2\sum_n\frac{G^2(n)m^2(n)}{Q^2+m^2(n)} + D_0 +D_1 Q^2\\
=\left[\sum_n 2 G^2(n) + D_0\right] -
Q^2 \left[\sum_n\frac{2 G^2(n)}{Q^2+m^2(n)} - D_1\right]\,.
\end{multline}
The first sum in the right-hand side of~(\ref{ren}) represents an
infinite constant and it must be renormalized in a chirally symmetric
way
\begin{equation}
\sum_n 2 G^2(n)+D_0 = \tilde{D}_0\,,
\end{equation}
where $\tilde{D}_0$ is different for the $S$ and $P$ channels.
Then we have
\begin{equation}
\int\frac{Q^2G^2(x)\,dx}{Q^2+m^2(x)}- \frac12 D_1Q^2=
\bar C Q^2\ln\!\left(\frac{Q^2+m^2(x)}{\mu^2}\right)+\dots\,.
\end{equation}
Repeating the discussion
in the previous section, which has led to Eq.~(\ref{conVA}),
one obtains in the scalar case
\begin{equation}
\label{co}
G^2_{S,P}(n)=a\left(\bar C +A_G^{S,P}e^{-B_Gn}\right), \qquad B_G>0,
\end{equation}
where
\begin{equation}
\bar C = \frac{3}{16\pi^2}\left(1+\frac{11\alpha_s}{3\pi}\right).
\end{equation}
Following the same arguments concerning the
parameter $B_m$, as in the previous section, we propose the
following parameterization of $S,P$ masses
\begin{equation}
\label{SPmasses}
m_{S,P}^2(n)=\bar{M}^2+an+A_m^{S,P}e^{-B_m n}.
\end{equation}

Now let us derive the asymptotic sum rules. In order to avoid
irrelevant infinite terms we consider the second derivative of
$S,P$-correlators. Making
use of~\eqref{lin}, \eqref{SPres}, \eqref{co}, and~\eqref{SPmasses}
we can write
\begin{equation}
\frac{d^2\Pi(Q^2)}{(dQ^2)^2}=4\sum_{n=0}^{\infty}
\frac{a\tilde{m}^2(n)\left(\bar{C}+A_Ge^{-B_Gn}\right)}{\left(Q^2+
\tilde{m}^2(n)+A_me^{-B_mn}\right)^3},
\end{equation}
where the notation $\tilde{m}^2 (n) \equiv \bar{M}^2 + a n$ 
for the linear part of the $S,P$  trajectory has been introduced.
Separating out the
first term in the sum and retaining only the leading order 
in exponential corrections, one obtains
\begin{multline}
\label{cors}
\frac14\frac{d^2\Pi(Q^2)}{(dQ^2)^2}\simeq
\frac{a(\bar C+A_G)}{\left(Q^2+\bar{M}^2+A_m\right)^3} \\
+\sum_{n=1}^{\infty}\left\{
\frac{a\bar{C}\tilde{m}^2(n)}{\left(Q^2+\tilde{m}^2(n)\right)^3}+
\frac{a\tilde{m}^2(n)A_Ge^{-B_Gn}}{\left(Q^2+\tilde{m}^2(n)\right)^3}-
\frac{3a\bar{C}\tilde{m}^2(n)A_me^{-B_mn}}{\left(Q^2+\tilde{m}^2(n)\right)^4}
\right\}
\end{multline}
In what follows the procedure is the same as for vector channels.

At this point one has to decide which
particles are chiral partners in order to guarantee
chirally symmetric results. One may think of two possibilities: a) the
$\pi$-meson belongs to the radial Regge trajectory being the parity-odd partner of
the lightest scalar meson, b) it does not belong to the radial Regge trajectory 
being an isolated Goldstone chiral particle and therefore the lightest scalar
meson is the parity-even partner of the $\pi'(1300)$ meson. In case a) 
the effective low-energy theory would be the 
linear $\sigma$-model, and in case b) it corresponds to the nonlinear
 $\sigma$-model. We are
going to check both variants. The corresponding sum rules are
presented in Appendix~B.

\section{Details of fits and results}

In order to compare the sum of the resonances with the OPE we have to choose
appropriate values for the inputs of the latter. 
They are the condensates $\langle\bar qq\rangle$ and
$\langle\left(G_{\mu\nu}^a\right)^2\rangle$, and $\alpha_s$.
On the resonance side we take as input the pion decay  constant $f_{\pi}$,
the pion pole residue
$Z_{\pi}=2\frac{\langle\bar qq\rangle^2}{f_{\pi}^2}$ (from
current algebra), and the slope $a$. 
The numerical values for all these parameters are taken at about the CSB scale 
$\sim 1$ GeV and presented in Appendix~F. In particular, the 
value\footnote{This number can be associated to the 
trade mark of "Brandy de Jerez 103, Osborne"}  of 
$f_{\pi} \approx 103$ MeV
is certainly different from its low-energy limit 
\cite{pich}  and corresponds to the matching of the OPE and the sums
of resonances at the latter scale (see \cite{yama} and references
therein).

Recall that in order to keep the discussion in simple terms and also to stay 
as close as possible to the chiral limit we consider here only non-strange mesons. So far we have also omitted any reference to the isospin degrees of freedom, but we have now to specify the particular pattern in which chiral symmetry is
restored in each channel.

In the $V,A$ channels we 
study isovector states, where the restoration of $U(1)_A$ symmetry should
be manifest in the large $N_c$ limit. We shall therefore
analyze the spectra of  $\rho$ and $a_1$ 
mesons and their radially excited relatives.
In this case five asymptotic sum rules contain
seven spectral parameters to be fitted:
$M$, $B_{m,F}$, $A_m^{V,A}$, and $A_F^{V,A}$. Thus, we need two additional 
physical inputs, which are chosen to be the masses of the ground states 
$\rho$ and $a_1$. Actually, we followed the best fit principle when taking
the  $a_1$ mass of about 1200 MeV in concordance with the masses of first 
excited vector and axial-vector states. We notice also that the  $a_1$ mass
is very sensitive to the variation of $f_{\pi}$, say, its decrease in 3 MeV
(up to 100 MeV) leads to diminishing  the  $a_1$ mass to 1180 MeV.

Let us comment on some numbers of Table~2 in Appendix~F. A
 possible
candidate for the state $m_V(2)$ could be the $\rho(1900)$
meson~\cite{pdg}. However, the width of this state is one order of
magnitude smaller than the widths of other vector states and it could 
correspond to a hybrid state (hybrids and glueballs do not lie on the
large-$N_c$ radial Regge trajectory for quarkonia). Thus, we do
not include this meson on the radial Regge trajectory.

Unfortunately, we cannot compare  the residues of
excited states with experiment. 
The possible exception is the vector channel due
to Eq.~\eqref{vres}. For example, our ansatz
predicts:
$\Gamma_{\rho(1450)\rightarrow e^+e^-}=2.9$ KeV. The relevant
widths $\Gamma_{V\rightarrow e^+e^-}$ are poorly known and not listed
in the Particle Data~\cite{pdg}. However, one may compare the results with
other independent model estimations. The corresponding numbers we have
found in~\cite{isg}:
$\Gamma_{\rho(1450)\rightarrow e^+e^-}=0.4$ KeV,
and in~\cite{henner}:
$\Gamma_{\rho(1450)\rightarrow e^+e^-}=3.5$ KeV,

Let us discuss now the D-wave vector mesons. Introducing these
states entails the appearance of three new parameters in the
asymptotic sum rules: $M_D$, $A_D$, and $B_D$. The first one can
be fixed by the mass of $\rho(1700)$-meson. One can also fix, say
$F_V(0)$ and $F_D(0)$. The average of existing estimates of electromagnetic
width for $\rho(1700)$-meson ($\Gamma_{\rho(1700)\rightarrow e^+e^-}=0.1$
KeV\cite{isg} and $\Gamma_{\rho(1700)\rightarrow e^+e^-}=2.7$
KeV\cite{henner}) presuppose\footnote{Strictly speaking, Eq.~\eqref{vres}
can be applied only to the S-wave vector mesons because of locality of vector
current in~\eqref{vme}. However, in the relativistic theory the situation
changes~\cite{isg} and transitions~\eqref{vme} for the D-wave vector states
and~\eqref{ame} for the axial-vector mesons (which are P-wave states) become
possible. Moreover, in the large-$N_c$ limit Eqs.~\eqref{vme}
and~\eqref{ame} are well defined even in the relativistic theory
since the resonances are narrow.
We use~\eqref{vres} for the D-wave vector mesons as a
rough estimate of coupling these states to the $e^+e^-$-annihilation.}
a rather large value for $F_D(0)$, of the order 60~MeV. However,
we could not find any reasonable solution for such large values of
residue of ground D-wave vector meson. Our estimations showed that
$F_D(0)$ (i.e., $A_D$) must be smaller by about two orders of
magnitude than  the value of  $F_D(0)$ obtained from the previous analysis. 
Then the contribution of D-wave vector mesons
to the physical quantities turns out to be below the
accuracy of large-$N_c$ counting. 

In the $S,P$ channel the experimental situation is a lot more confusing. In the
large $N_c$ limit one might 
equally look for the restoration of
$U(1)_A$ symmetry relating isotriplet pseudoscalar and scalar states,
i.e. pions ($I=1$) and $a_0$ ($I=1$) mesons and their radial excitations.
On the other hand, one could examine the restoration $SU(2)_L \times SU(2)_R$
symmetry relating pions ($I=1$) and $f_0$ ($I=0$) in ground and excited states.
Both scenarios are compatible if one assumes that there is a
degeneracy between $a_0$ and $f_0$ states. To some extent
the latter isoscalar
 \cite{olos}  
can be related to $f_0 (980)$ from the PDG data \cite{pdg}. But for
the former isotriplet there is no firm identification with  $a_0(980)$ 
from the PDG data \cite{pdg}.
Rather the unitarized fits of pion scattering data \cite{olos,pel} seem to
indicate the dynamical origin  of $a_0(980)$ as a meson bound
state.
Thus in order to deal with well established states in the $S,P$ channels
in the present 
paper we have restricted our analysis to asymtotic sum rules 
relating the isovector pseudoscalar channel (i.e. pions) to the
isoscalar meson channel (i.e. $f_0$ states). One should of course bear in mind that including strange quarks and moving from $SU(2)$ to $SU(3)$ could bring in 
some relevant changes in the scalar sector. 
  
In this sector we have considered two possibilities, namely the cases a) and b) 
at the end of the previous section. Case a) is labeled "$\pi$-in". By this we mean that
the pion as well as the lightest scalar meson are taken to be Regge states. We
have here seven spectral parameters: $\bar M$, $B_{m,G}$, $A_m^{S,P}$,
and $A_G^{S,P}$, which are subject to three asymptotic sum rules. As was
pointed above, $B_m$ is universal for all channels. The three additional
conditions are chosen to be: $m_S(0) = 1$ GeV , $m_P(0)= 0$, and $m_P(1) = 1.3$ GeV.
This presupposes that the pion and the above scalar are chiral partners. Popular linear 
$\sigma$-models
often require a much lighter scalar with a mass around $600$ MeV. However our fit favors
a heavier scalar quarkonium  to provide a realistic value
\footnote{ When lowering the input mass of the scalar meson to $600$ MeV
one gets  $L_8 = 2.8\cdot 10^{-3}$, much higher than the phenomenological 
estimate  $L_8 = (0.8\pm0.3)\cdot 10^{-3}$ \cite{pich}} 
of $L_8$, in agreement with the conclusions of
\cite{olos,pel}. The numerical results are presented in Appendix F.

In case b) the $\pi$-meson is assumed 
not to belong to the radial Regge trajectory being a Goldstone boson that is actually
decoupled from the radial Regge trajectory in the CSR limit. 
This case is labeled "$\pi$-out". According to our fits (the lightest 
scalar state is again assumed to have a mass of
about 1~GeV), the main qualitative
difference between the linear and nonlinear cases is that the first one
predicts a scalar meson with a mass of about 1.44~GeV (an iso-singlet resonance with  a mass of
 1.5~GeV exists but it is widely believed to be a glueball), while the
second one does not.
%%%%%%%%%%%%%%

We emphasize again that our analysis in the $S,P$ channels has been performed for $SU(2)$
multiplets including iso-triplet pesudoscalar and iso-singlet scalar
mesons.
We have also neglected  current quark masses adopting the chiral
limit. The situation may be drastically different for isotriplet
scalars and  for $SU(3)$
multiplets.
In particular, the  $a_0(980)$ (isotriplet) mesons may well be  dynamical
resonances which decouple in the large $N_c$ limit whereas the mesons
$a_0(1450)$ may be dominantly\footnote{We are grateful to the referee who has drawn
our attention to this possibility.} ground quarkonium states \cite{cirig}.
Their rather large masses could arise owing to a strong mixing
between a lighter dynamical scalar  $a_0$ and a heavier  $a'_0$
quarkonium state with masses of order $1.2$ GeV \cite{schech, cirig}. 
We plan to investigate this scenario with the help of asymptotic sum
rules in a forthcoming paper, taking also 
into account light quark mass effects.
%%%%%%%%%%%%%%%%%%%%%%%

The quantities $G_P(n)$ are related to the corresponding weak
decay constants $F_P(n)$ through the relation
$$
F_P(n)=\frac{2m_qG_P(n)}{m_P(n)}.
$$
In particular, accepting the average value $m_q$ for the current
masses of $u$- and $d$-quarks to be equal to 6~MeV, we have the
following estimates for the~$F_{\pi(1300)}$: $F_P(1)=1.7$~MeV ("$\pi$-in")
and $F_P(0)=1.9$~MeV ("$\pi$-out"). Both values are consistent 
with some previous 
theoretical predictions \cite{domin,aam},  smaller than
the estimates from the Finite-Energy Sum Rules \cite{kat2} and
larger than the predictions from the non-local quark model\cite{volkov}.

It should be also noticed that, allowing for a 10\%-accuracy,
the first two chiral pairs of resonances saturate the chiral
constant $L_{10}$ introduced in \cite{gl} almost completely
\footnote{The analytical formulas for the chiral constants $L_8, L_{10}$ 
as well as for $\Delta m_{\pi}$ are given in Appendix E}. But this is not
the case for the quantity $\Delta m_{\pi}$: here, for our fits, one needs to
retain about seven pairs. The reason is that the value of this quantity
is very sensitive to the violation of the asymptotic sum rules with gluon
condensate (strictly speaking, the difference of these sum rules): if one
likes to calculate the contribution of the first $N$ pairs of $V,A$
resonances to $\Delta m_{\pi}$, one has then to deal with
exactly this number of resonances in the asymptotic sum rules from the
very beginning~\cite{a}.

\section{Summary}

In the present work we have considered the matching of the
vector, axial-vector, scalar, and pseudoscalar meson mass spectra
$m^2(n)$ ($n$ is the radial quantum number) from Regge theory
with universal slope to the Operator Product Expansion of quark currents. The analysis has been
carried out for the light non-strange mesons in the large-$N_c$
and chiral limits.  Let us summarize the important
lessons that we gained from our analysis.
\begin{itemize}
\item  The matching to the OPE cannot be achieved by a simple linear
parameterization of the mass spectrum, the linear trajectory ansatz. 
\item  The convergence of the generalized Weinberg sum rules requires
the universality of slopes and intercepts for parity conjugated trajectories.
%The tree-level limit implies universality of intercepts for all
%V\!,A,S,P spectra.
\item There must exist deviations from the
linear trajectory ansatz triggered by  chiral
symmetry breaking. These deviations must decrease at least
exponentially with $n$.
\item
There are also deviations from constant
residues (decay constants) $F^2(n)$
(or for the quantities $G^2(n)$ in the scalar case). The
analytic structure of OPE imposes again an  exponential decrease on
these deviations (or faster).
\item
For heavy states, the D-wave vector mesons have to decouple from
asymptotic sum rules. This fact implies the exponential (or faster)
decreasing the corresponding decay constants $F^2_D(n)$.
\item Our results seem to exclude a light $\sigma(600)$ particle as a quarkonium state and
rather favor the non-linear realization of chiral symmetry with the lightest 
scalar of  mass $\sim 1$ GeV, its chiral partner being the $\pi'(1300)$.
\item As a consequence of our approach the quantities $L_8, L_{10}$ and $\Delta m_\pi$
are obtained, in  satisfactory agreement with the phenomenology.
 \end{itemize}

Unfortunately, the underlying dynamics, which generates the
non-linear contributions to the spectra of meson masses and
residues, is not well known. We can only say that these deviations from
the string picture seem to parameterize the chiral symmetry breaking
in QCD~\cite{AABE}
and, hence, must be proportional to powers of the chiral condensate
$\langle\bar qq\rangle$\cite{PhLB}. Developing a theory of these non-linear
contributions is an interesting task for future.

Another interesting problem to be examined concerns the $SU(3)$ extension with
inclusion of current quark masses into consideration and a possible
resolution
of the $a_0$ meson ambiguity \cite{cirig}. Finally, one could get
some insight by overlapping the present, direct-resonance
approach
with spectral density methods \cite{groote}.

\acknowledgments

The work of S.A., A.A. and V.A. was supported by
 the Program "Universities of Russia:
Basic Research" (Grant 02.01.016). The work of D.E. was supported by
the EURIDICE Network, grant FPA-2001-3598 and grant 2001SGR-00065.
All authors enjoyed the support
of INTAS-2000 Project 587. A.A. and D.E. wish to thank the warm
hospitality of the Pontificia Universidad Catolica de Chile. We are also
grateful to A. Pineda  and S. Peris for useful discussions.

\appendix
\section{Note on the role of dimension-two condensate}
Let us examine the possibility of having a dimension-two gluon condensate and
show its unimportance for fitting meson parameters, 
at least, in the large-$N_c$ limit. 
The dimension-two gluon condensate $\lambda^2 < 0$ 
("tachyonic gluon mass") was
introduced in~\cite{z1}. This dimension-two condensate cannot be characterized 
by a local
gauge-invariant operator: for
speculations concerning origin, measurement, and physical meaning
of this condensate see~\cite{z1,z2,kondo}.

In~\cite{z1,z2} the following relevant modification of the OPE for quark currents
was proposed
\begin{equation}
\label{delVA}
\Delta\Pi^V(Q^2)=
\Delta\Pi^A(Q^2)=-\frac{\alpha_s}{4\pi^3}\cdot\frac{\lambda^2}{Q^2}
\end{equation}
\begin{equation}
\label{delSP}
\Delta\Pi^S(Q^2)= \Delta\Pi^P(Q^2)= -\frac{3\alpha_s}{2\pi^3}\lambda^2 
\ln\!\frac{\Lambda^2}{Q^2}.
\end{equation}
There is no problem to derive such terms from the resonance sums in the
$V,A$ channels: their introduction is compatible with the 
ans\"atze~\eqref{conVA},~\eqref{VAmasses} (see the corresponding sum rules
in the next Appendix, 
Eqs.~\eqref{qfourv},~\eqref{qfoura}). 
However, in order to reproduce the pertinent asymptotical 
terms in the scalar channels one has to essentially modify the linear
part of the ansatz ~\eqref{SPres} for the residues of the $S,P$ resonances.
We consider two types of possible contributions: an asymptotically constant
shift and one affecting only the physical pion residue ($n=0$), 
\begin{equation}
\label{SPlambda}
Z_{S,P}(n)\longrightarrow Z^\lambda_{S,P}(n)\equiv 
2G_{S,P}^2(n)m_{S,P}^2(n) + G^\lambda_0\frac{dm_{S,P}^2(n)}{dn} + 
\tilde{Z}_\pi \ \delta_{n,0}.
\end{equation}
For this ansatz 
the appropriate resonance sums~\eqref{ren} generate the contribution
\begin{equation}
\Delta\Pi^{S,P} \simeq G^\lambda_0 
\ln\!\left(\frac{Q^2+m^2(x)}{\mu^2}\right)+\dots\, ,
\end{equation}
which saturates the dimension-two asymptotics~\eqref{delSP} if
\begin{equation}
\label{Glambda}
G^\lambda_0=-\frac{3\alpha_s}{2\pi^3}\lambda^2.
\end{equation}
Then some changes are to be done in the sum rules of the next Appendix.

On the other hand the modification~\eqref{SPlambda} evidently
affects the pion pole residue $Z_\pi$ (at $n = 0$)
\begin{equation}
\label{zpilambda}
Z_\pi \equiv 2\frac{{\langle\bar qq\rangle}^2}{f_{\pi}^2}
= \tilde{Z}_\pi + a G^\lambda_0. 
\end{equation}
If we put  $\tilde{Z}_\pi = 0$ 
and the pion is put on the radial Regge trajectory 
the required value of the "gluon mass"
should be $\lambda^2 \simeq - 2\, \mbox{\rm GeV}^2$.  
However this value is, at least, 
one order of magnitude higher that any known theoretical estimations \cite{z2},
$\lambda^2 = - (0.2 \div 0.5) \mbox{\rm GeV}^2$, 
and phenomenological bounds (from the
analysis of $\tau$-lepton decay) 
\cite{domsch}, $\lambda^2 = - (0.05 \pm 0.08) \mbox{\rm GeV}^2$.

We conclude that for realistic values  of the dimension-two condensate 
$\lambda^2$ its contribution to the residues Eqs.~\eqref{SPlambda},
\eqref{zpilambda} is negligible. Furthermore one can check that it is 
certainly less than 5\% for meson masses and decay constants. Still a good
open question is about what is a physical observable which is sufficiently 
sensitive to its presence.

\section{Sum rules}

After all summations in~\eqref{cor2} one arrives at the expansion
\begin{equation}
-\frac{1}{2}\frac{d\Pi^{V,A}(Q^2)}{dQ^2}\simeq
\sum_{k=1}^{\infty}\frac{c_k^{V,A}}{Q^{2k}}.
\label{VAOPE}
\end{equation}
Substituting the expressions for $c_k$ and comparing~\eqref{VAOPE}
with OPE~\eqref{V} and~\eqref{A}
we obtain the following asymptotic sum rules in the $V,A$-channels.

\noindent
At $1/Q^2$
\begin{equation}
c_1^{V,A}=\frac{N_c}{24\pi^2}\left(1+\frac{\alpha_s}{\pi}\right)=C.
\end{equation}
At $1/Q^4$ (the dimension two condensate is taken to be 
zero as implied
by the sum rules in the $S,P$ channels)
\begin{equation}
\label{qfourv}
c_2^V=a(C+A_F^V)-C\left(\frac{1}{2}a+M^2\right)+A_F^V\Delta^{(1)}_F=0,
\end{equation}
\begin{equation}
\label{qfoura}
c_2^A=a(C+A_F^A)-C\left(\frac{1}{2}a+M^2\right)+A_F^A\Delta^{(1)}_F=
-f_{\pi}^2.
\end{equation}
At $1/Q^6$
\begin{multline}
\label{gv}
c_3^A=-2a(C+A_F^V)(M^2+A_m^V)+C\left(M^2\left(M^2+a\right)+\frac16a^2\right)-
2CA_m^V\Delta^{(1)}_m-2A_F^V\Delta^{(2)}_F \\
=\frac{\alpha_s}{12\pi}\langle\left(G_{\mu\nu}^a\right)^2\rangle,
\end{multline}
\begin{multline}
\label{ga}
c_3^A=-2a(C+A_F^A)(M^2+A_m^A)+C\left(M^2\left(M^2+a\right)+\frac16a^2\right)-
2CA_m^A\Delta^{(1)}_m-2A_F^A\Delta^{(2)}_F \\
=\frac{\alpha_s}{12\pi}\langle\left(G_{\mu\nu}^a\right)^2\rangle.
\end{multline}
At $1/Q^8$ in $\Pi^V(Q^2)-\Pi^A(Q^2)$
\begin{multline}
\label{last}
c_4^V-c_4^A=3a(C+A_F^V)(M^2+A_m^V)^2-3a(C+A_F^V)(M^2+A_m^V)^2+
6C(A_m^V-A_m^A)\Delta^{(2)}_m \\
+3(A_F^V-A_F^A)\Delta^{(3)}_F = -12\pi\alpha_s\langle\bar qq\rangle^2.
\end{multline}
In Eqs.~\eqref{qfourv}-\eqref{last} the following notations were
introduced (the symbol $i$ denotes $m$ or $F$ or $G$)
\begin{equation}
\label{delt1}
\Delta^{(1)}_i=\frac{a}{e^{B_i}-1},
\end{equation}
\begin{equation}
\Delta^{(2)}_i\equiv 
\Delta^{(2)}_i(M) =\frac{a(-M^2+(M^2+a)e^{B_i})}{(e^{B_i}-1)^2},
\end{equation}
\begin{equation}
\label{delt3}
\Delta^{(3)}_i\equiv 
\Delta^{(3)}_i(M)=\frac{a\left[-a(a+2M^2)+ae^{B_i}(3a+2M^2)+
(M^2+a)^2(e^{B_i}-1)^2\right]}{(e^{B_i}-1)^3}.
\end{equation}

In the scalar case we can write down analogous to~\eqref{VAOPE}
expansion:
\begin{equation}
\label{SPOPE}
\frac14\frac{d^2\Pi^{S,P}(Q^2)}{(dQ^2)^2}\simeq
\sum_{k=1}^{\infty}\frac{c_k^{S,P}}{Q^{2k}}.
\end{equation}
Comparing~\eqref{SPOPE} with OPE~\eqref{S} and~\eqref{P} one obtains
the asymptotic sum rules. We write them for the $\pi$-out case.
 The $\pi$-in case can be easily obtained too.

\noindent
At $1/Q^2$
\begin{equation}
c_1^{S,P}=\frac{N_c}{32\pi^2}\left(1+\frac{11\alpha_s}{3\pi}\right)=
\frac{\bar C}{2}.
\end{equation}
At $1/Q^6$
\begin{multline}
\label{1s}
c_2^S=a(\bar C+A_G^S)(\bar{M}^2+A_m^S)-
\frac{\bar C}{2}\left(\bar{M}^2\left(\bar{M}^2+a\right)+\frac16a^2\right)+
A_G^S\Delta^{(2)}_G(\bar{M})\\
= \frac{\alpha_s}{16\pi}\langle\left(G_{\mu\nu}^a\right)^2\rangle,
\end{multline}
\begin{multline}
\label{1p}
c_2^P=\frac{\langle\bar qq\rangle^2}{f_{\pi}^2}+
a(\bar C+A_G^P)(\bar{M}^2+A_m^P)-
\frac{\bar C}{2}\left(\bar{M}^2\left(\bar{M}^2+a\right)
+\frac16a^2\right)+A_G^P\Delta^{(2)}_G(\bar{M})\\
= \frac{\alpha_s}{16\pi}\langle\left(G_{\mu\nu}^a\right)^2\rangle.
\end{multline}
At $1/Q^8$ in the $\Pi_S(Q^2)-\Pi_P(Q^2)$
\begin{multline}
c_3^S-c_3^P=-3a(\bar C+A_G^S)(\bar{M}^2+A_m^S)^2+3a(\bar C+A_G^P)(\bar{M}^2+A_m^P)^2
-3\bar C(A_m^S-A_m^P)\Delta^{(2)}_m(\bar{M}) \\
-3(A_G^S-A_G^P)\Delta^{(3)}_G(\bar{M}) = -18\pi\alpha_s\langle\bar qq\rangle^2.
\end{multline}

\section{Reference formulae}

The quantities $\Delta_i^{(j)}$ ($j=1,2,3$, see~\eqref{delt1}-\eqref{delt3})
appear due to the sums ($B>0$)
$$
\sum_{n=1}^{\infty}e^{-Bn}=\frac{1}{e^B-1},
$$
$$
\sum_{n=1}^{\infty}e^{-Bn}n=\frac{e^B}{(e^B-1)^2},
$$
$$
\sum_{n=1}^{\infty}e^{-Bn}n^2=\frac{e^B(e^B+1)}{(e^B-1)^3}.
$$

The summation of the linear in $n$ part is carried out by virtue of the
following asymptotic representations (to be precise,once we separated out the first
state, one has to make the shift $M^2\rightarrow M^2+a$ in the
expressions below)
\begin{multline}
\label{a1}
\sum_{n=0}^{\infty}\frac{1}{(Q^2+M^2+an)^2}=
\frac{1}{a^2}\psi\left(1,\frac{Q^2+M^2}{a}\right)=
\frac{1}{a}\left\{\frac{1}{Q^2}-\frac{1}{Q^4}\left(M^2-\frac12a\right)
\right.+\\
\left.\frac{1}{Q^6}\left(M^4-aM^2+\frac16a^2\right)-
\frac{M^2}{Q^8}\left(M^2-\frac12a\right)\left(M^2-a\right)\right\}+
\mathcal{O}\left(\frac{1}{Q^{10}}\right),
\end{multline}
\begin{equation}
\label{a2}
\sum_{n=0}^{\infty}\frac{M^2+an}{(Q^2+M^2+an)^3}=
\frac{Q^2}{2a^3}\psi\left(2,\frac{Q^2+M^2}{a}\right)+
\frac{1}{a^2}\psi\left(1,\frac{Q^2+M^2}{a}\right),
\end{equation}
where
\begin{multline}
\label{a3}
\frac{Q^2}{2a^3}\psi\left(2,\frac{Q^2+M^2}{a}\right)=
\frac{1}{a}\left\{-\frac{1}{2Q^2}+\frac{1}{Q^4}\left(M^2-\frac12a\right)
\right.-\\
\left.\frac{3}{2Q^6}\!\left(\!M^4-aM^2+\frac16a^2\!\right)\!+\!
\frac{2M^2}{Q^8}\!\left(\!M^2-\frac12a\!\right)\left(\!M^2-a\!\right)
\right\}\!+\!\mathcal{O}\!\left(\!\frac{1}{Q^{10}}\!\right)\!\!.
\end{multline}
Summing Eq.~\eqref{a1} and Eq.~\eqref{a3} in Eq.~\eqref{a2},
one can see the absence of term at $1/Q^4$ in the $S,P$-correlators.
The corrections contribute only to the following terms, beginning
with $1/Q^6$. Thus, in the scalar case the asymptotics corresponding to a
dimension two condensate vanishes identically and   we do not have 
a relevant sum rule.

\section{D-wave vector mesons}

The string model does not take into account the spin of the quarks. 
This introduces two type of states depending on the relative angular
momentum. In short, there exist S-wave and
D-wave $\rho$-mesons and this results in doubling of the $\rho$-meson
trajectory~\cite{ani,ani2}. Although conceptually simple, this
complicates the analysis considerably.

The two kinds of
$\rho$-mesons should enter the sum rules independently. This observation,
generally speaking, leads to the constrain in Eq.~(\ref{sr1})
\begin{equation}
\frac{F_{V}^2(n)}{a_{V}}+\frac{F_{D}^2(n)}{a_{D}}
=\frac{F_A^2(n)}{a_A},\qquad n\rightarrow \infty ,
\end{equation}
where the label $D$ stands for the D-wave $\rho$-mesons.
It is seen from numerical estimations (especially for the
correct ansatz $a_{V}=a_{D}=a$) that $F_{D}(n)$ must be very
small compared with $F_{V}(n)$.
This follows also from consideration of the matrix
element~$\langle0|\bar{q}\gamma_{\mu}q|\rho^0\rangle$. It is clear
qualitatively that the D-wave final state is strongly suppressed
compared with the S-one, although a concrete numerical answer will
depend on how one models the wave function of $\rho^0$-meson.
In the non-relativistic case this statement is trivial: S-wave
component of wave function behaves as a constant and D-one does as $r^2$
at small distances $r$. Hence the latter
tends to zero since annihilation is a point-like process.
Thus, D-wave $\rho$-mesons, at least asymptotically, should drop out
from the
sum rules in contrast to what has been stated in~\cite{sim}, 
where this doubling
was examined by means of quasi-classical string analysis.

Let us consider the linear ansatz for the mass spectrum of
D-wave vector mesons
\begin{equation}
\label{D-mass}
m_D^2(n)=M_D^2+an.
\end{equation}
Direct insertion of this ansatz into sum rules~\eqref{bea} would
automatically imply by the CSR that $m_D=M$. This contradicts
phenomenology~\cite{ani,ani2}: S- and D-trajectories
seem to have a
constant splitting at any energy. In order to satisfy the sum
rules of type~\eqref{bea}, the D-wave residues should decrease at
least exponentially. We propose the following parameterization
\begin{equation}
F_D^2(n)=a A_D e^{-B_Dn},\qquad B_D>0.
\end{equation}
As these residues are exponentially small we do not consider the
exponential corrections to the mass spectrum of D-wave
$\rho$-mesons~\eqref{D-mass} in our approximation.

There arises the
question as to how the chiral limit affects the behavior of the D-wave
vector trajectory. One may think of two possibilities: i) the D-wave mesons
approach the S-wave vector trajectory, implying
asymptotic degeneration; ii) D-wave
$\rho$-mesons decouple. The latter variant seems to be more plausible since
the former one signifies the doubling of states on the vector
trajectory compared with the axial-vector one. In addition, the
decoupling must take place in any case by virtue of the
qualitative arguments mentioned above.

\section{Electromagnetic pion mass difference $\Delta m_{\pi}$
and chiral constants $L_8$, $L_{10}$.}

Given an ansatz for mass spectrum $m_J^2(n)$ and decay constants $F_{J}^2$,
one can calculate the electromagnetic pion mass difference
$\Delta m_{\pi}\equiv m_{\pi^+}-m_{\pi^0}$, the chiral constant
$L_{10}$~\cite{gl} (parameterizing the decay $\pi\rightarrow e\nu\gamma$),
and a $K\rightarrow\pi$ matrix element of the electromagnetic penguin
operator $Q_7^{3/2}$. An example of such a calculation for the LTA was
provided in~\cite{p1}.
Besides, in the scalar sector we can
calculate the chiral constant $L_8$~\cite{gl} (parameterizing the ratio of
current quark masses). We did not consider the operator $Q_7^{3/2}$
since this quantity strongly depends on the cutoff and is
of a limited phenomenological interest. The other quantities
are determined by
\begin{equation}
\label{L10}
L_{10}=-\frac18\frac{d}{dQ^2}\left[Q^2\left(\Pi^V(Q^2)-
\Pi^A(Q^2)\right)\right]_{Q^2=0}\,,
\end{equation}
\begin{equation}
\label{dmp}
\Delta m_{\pi}=-\frac{3\alpha}{16\pi m_{\pi}
f_{\pi}^2}\int_0^{\infty}dQ^2Q^2\left(\Pi^V(Q^2)-
\Pi^A(Q^2)\right),
\end{equation}
\begin{equation}
\label{L8}
L_8=\frac{f_{\pi}^4}{32\langle\bar qq\rangle^2}\frac{d}{dQ^2}
\left[Q^2\left(\Pi^S(Q^2)-\Pi^P(Q^2)\right)\right]_{Q^2=0}\,.
\end{equation}
In our approach, one can directly sum over all resonances in this
quantities due to the exact cancellation of divergent parts. Making
use of notations \eqref{VAmasses},\eqref{lin} and \eqref{D-mass}
one can write for~\eqref{L10}-\eqref{L8} the convergent sums (we pick out
the ground states, treat them exactly  and  retain only 
the leading and linear in exponential terms
for the other ones)
\begin{multline}
\label{L10b}
L_{10}=\frac{a}{4}\left\{\frac{C+A_F^A}{M^2+A_m^A}-\frac{C+A_F^V}{M^2+A_m^V}
\right.\\
\left.+\sum_{n=1}^{\infty}
\frac{\bar{m}^2(n)e^{-B_Fn}\left(A_F^A-A_F^V\right)-
Ce^{-B_mn}\left(A_m^A-A_m^V\right)}
{\bar{m}^4(n)}\right\},
\end{multline}
\begin{multline}
\label{dmpb}
\Delta m_{\pi}=\frac{3\alpha a}{16\pi m_{\pi}f_{\pi}^2}\left\{
(C+A_F^A)(M^2+A_m^A)\ln\!\frac{M^2+A_m^A}{\mu^2}\right. \\
-(C+A_F^V)(M^2+A_m^V)\ln\!\frac{M^2+A_m^V}{\mu^2}
+\sum_{n=1}^{\infty}\left(C\bar{m}^2(n)\ln\!\frac{m_A^2(n)}{m_V^2(n)}\right.\\
\left.\left.+\left[Ce^{-B_mn}\left(A_m^A-A_m^V\right)+
\bar{m}^2(n)e^{-B_Fn}\left(A_F^A-A_F^V\right)\right]
\ln\!\frac{\bar{m}^2(n)}{\mu^2}\right)\right\}.
\end{multline}
In the $\pi$-in case
\begin{multline}
\label{L8b} L_{8}=\frac{f_{\pi}^4a}{16\langle\bar
qq\rangle^2}\left\{\bar C+A_G^S+
\sum_{n=1}^{\infty} e^{-B_Gn}\left(A_G^S-A_G^P\right)\right\}=\\
\frac{f_{\pi}^4a}{16\langle\bar qq\rangle^2}\left\{\bar C+A_G^S+
\frac{\left(A_G^S-A_G^P\right)}{e^{B_G}-1}\right\},
\end{multline}
while for the $\pi$-out case
\begin{equation}
\label{L8b2} L_{8}=\frac{f_{\pi}^4a}{16\langle\bar
qq\rangle^2}
\sum_{n=0}^{\infty} e^{-B_Gn}\left(A_G^S-A_G^P\right)=
\frac{f_{\pi}^4a}{16\langle\bar qq\rangle^2}
\frac{\left(A_G^S-A_G^P\right)}{1-e^{-B_G}}.
\end{equation}
The parameter $\mu$ in Eq.~\eqref{dmpb} is a normalization
scale. The result does not depend on it as it is seen from the
difference of sum rules \eqref{gv} and \eqref{ga}.

Accepted estimates for $L_{10}$ and $L_8$ from phenomenology are:\\
$L_{10}|_{\mbox{\scriptsize phen}}=(-5.5\pm0.7)\cdot10^{-3}$\cite{pich}
and $L_8|_{\mbox{\scriptsize phen}}=(0.8\pm0.3)\cdot10^{-3}$\cite{pich}.

\section{Numerical analysis}

In this Appendix we give an example of the meson mass spectra
resulting from our work. The inputs general for all tables (if any)
are: $a=(1120\,\mbox{MeV})^2$, $\langle\bar
qq\rangle=-(240\,\mbox{MeV})^3$,
$\frac{\alpha_s}{\pi}\langle\left(G_{\mu\nu}^a\right)^2\rangle=
(360\,\mbox{MeV})^4$, $f_{\pi}=103\,\mbox{MeV}$,
$Z_{\pi}=2\frac{\langle\bar qq\rangle^2}{f_{\pi}^2}$,
$\alpha_s=0.3$. The units are: $m(n)$, $F(n)$, $G(n)$ --- MeV; $A_{m}$
--- MeV${}^2$; $A_F$, $A_G$, $B_{F,G,m}$ --- MeV${}^{0}$.

\begin{center}
\TABLE{
\caption{An example of parameters for the mass spectra of our
work. The corresponding experimental values~\cite{ani2,pdg} (if
any) are displayed in brackets.}
\vspace{3mm}
\begin{tabular}{|c||l|l|}
  \hline
  Case & \hspace{18mm} Inputs & \hspace{22mm} Fits and constants \\
  \hline
  \hline
   $VA$ & \begin{tabular}{l}
   $m_V(0)=770\,(769.3\pm0.8)$,\\
   $m_A(0)=1200\,(1230\pm40)$,\\
   \end{tabular}
   & \begin{tabular}{l}
   $M=920$, $B_m=0.97$, $B_F=0.72$,\\
   $A_m^V=-500^2$, $A_m^A=770^2$,\\ $A_F^V=0.0012$,
   $A_F^A=-0.0031$,\\ $L_{10}=-6.5\cdot10^{-3}$, $\Delta m_{\pi}=2.3$
   \end{tabular} \\
  \hline
  \hline
  \begin{tabular}{c}
   $SP$\\
   ($\pi$-in)
   \end{tabular}
   & \begin{tabular}{l}
   $m_S(0)=1000$,\\
   $m_P(0)=0$,\\
   $m_P(1)=1300\,(1300\pm100)$,\\
   $B_m=0.97$
   \end{tabular}
   & \begin{tabular}{l}
   $\bar{M}=840$, $B_G=0.42$,\\ $A_m^S=550^2$, 
   $A_m^P=-840^2$, \\$A_G^S=-0.0009$, $A_G^P=0.0004$, \\
   $L_8=1.0\cdot10^{-3}$
   \end{tabular} \\
  \hline
  \hline
   \begin{tabular}{c}
   $SP$\\
   ($\pi$-out)
   \end{tabular}
   & \begin{tabular}{l}
   $m_S(0)=1000$,\\
   $m_P(0)=1300\,(1300\pm100)$,\\
   $m_P(1)=1800\,(1801\pm13)$,\\
   $B_m=0.97$
   \end{tabular}
   & \begin{tabular}{l}
   $\bar{M}=1470$, $B_G=1.27$,\\ $A_m^S=-1080^2$,
   $A_m^P=-690^2$,\\ $A_G^S=0.0213$, $A_G^P=0.0067$,\\
   $L_8=0.9\cdot10^{-3}$
   \end{tabular} \\
  \hline
\end{tabular}}

\TABLE{
\caption{Mass spectrum and residues for the parameter sets of
Table~1. The known experimental values~\cite{ani2,pdg} are
displayed in brackets. Not all the masses of scalar mesons 
are related to experimental ones since their
correspondence is not well established yet because of strong 
mixing and unitarization effects. 
When comparing the predicted masses to the physical
mass spectrum no attempt has been made to correct for the non-zero quark mass other than
for the pion. Except for the pion the relative effect of quark masses is very small.}
\vspace{3mm}
\begin{tabular}{|c||c|l|l|l|l|}
  \hline
  \!$n$\! & \!out\! & $\hspace{12.5mm} 0$ & $\hspace{14mm} 1$ & $ \hspace{14mm}2$ &
  $\hspace{14mm} 3$ \\
  \hline
  \hline
 \!$m_V(n)$\!  & & $770\,(769.3\!\pm\!0.8)$ & $1420\,(1465\pm25)$ 
& $1820$ &
  $2140\,(2149\pm17)\!\!$ \\
  \!$F_V(n)$\! &  & $138\,(154\!\pm\!8)$ & $135$ &  $133$  & $133$ \\
  \hline
  \!$m_A(n)$\! & & $1200\,(1230\!\pm\!40)$& $1520\,(1640\pm40)$ &
  $1850\,(1971\pm15)$ & $2150\,(2270\pm50)\!\!$\\
  \!$F_A(n)$\! &  & $116\, (123\!\pm\!25)$ & $125$ & $128$ &  $130$  \\
  \hline
  \hline
  \!$m_S(n)$\! &  & $1000 \,(980\!\pm\!10)$ & $1440$ & $1800\, (1713\!\pm\!6)$ 
& $2100$ \\
  \!$G_S(n)$\! &  & $176$ & $178$ & $178$ & $179$ \\
  \hline
  $m_P(n)$ &  & $0$ & $1300\,(1300\pm100)\!\!$ &$1760\,(1801\pm13)$ &
  $2100\,(2070\pm35)\!\!$ \\
  $G_P(n)$ &  & -- & $179$ & $179$ & $179$  \\
  \hline
  \hline
  \!$m_S(n)$\! &  & $1000\, (980\!\pm\!10)$ & $1730\, (1713\!\pm\!6)$ 
& $2120$ & $2420$ \\
  \!$G_S(n)$\! &  & $243$ & $199$ & $185$ & $181$ \\
  \hline
  \!$m_P(n)$\! & 0 & $1300\,(1300\pm100)\!\!$\! &$1800\,(1801\pm13)$ &
  $2150\,(2070\pm35)$\!\! & $2430\,(2360\pm30)\!\!$\\
  \!$G_P(n)$\! & -- & $201$ & $186$ & $181$ & $180$  \\
  \hline
\end{tabular}}
\end{center}

\end{document}